\documentstyle[12pt]{article}
\begin{document}
\title{An Analog Analogue of a Digital Quantum Computation\thanks{This
work was supported in part by The Department of Energy under cooperative agreement
DE-FC02-94ER40818}}
\author{Edward Farhi\thanks{farhi@mitlns.mit.edu} \\ Center for Theoretical
Physics
\\ Massachusetts Institute of Technology \\ Cambridge, MA  02139 \\[2ex] Sam
Gutmann\thanks{sgutm@nuhub.neu.edu} \\ Department of Mathematics \\ Northeastern
University \\ Boston, MA 02115 \\[3ex]
{\normalsize MIT-CTP-2593,~~quant-ph/9612026 \qquad \qquad December 1996}}
\date{}
\maketitle
\vspace{-.4in}
\begin{abstract}
We solve a problem, which while not fitting into the usual paradigm, can be viewed as a
quantum computation.  Suppose we are given a quantum system described by an
$N$ dimensional Hilbert space with a Hamiltonian of the form $E |w\rangle\langle w|$
where $| w\rangle$ is an unknown (normalized) state. We show how to discover $|
w\rangle$ by adding a Hamiltonian (independent of $| w\rangle$) and evolving for a
time proportional to $N^{1/2}/E$.  We show that this time is optimally short.  This
process is an analog analogue to Grover's algorithm, a computation on a conventional (!)
quantum computer which locates a marked item from an unsorted list of $N$ items in a
number of steps proportional to $N^{1/2}$.
\end{abstract}

\newpage

 Although a quantum computer, beyond certain elementary gates, has not yet
been constructed, a paradigm  \cite{bib:1}  for quantum computation is in
place.  A
quantum computer is envisaged as acting on a collection of spin 1/2
particles sitting at
specified sites.  Each elementary operation is a unitary transformation
which acts on the
spins at one or two sites.  A quantum computer program, or algorithm, is a
definite sequence of such unitary transformations.  For a given initial
spin state, the
output of the program is the spin state after the sequence of
transformations has acted.
The length of the algorithm is equal to the number of elementary unitary
transformations which make up the algorithm.

This framework for quantum computation is general enough that any ordinary
digital
computer program can be turned into a quantum computer algorithm.  (It is
required
that the ordinary program be reversible; however any ordinary computer
program can
be written in reversible code.)  Quantum computers can go beyond ordinary
computers when they act on superpositions of states and take advantage of
interference
effects.  An example of a quantum algorithm which outperforms any classical
algorithm
designed to solve the same problem is the Grover algorithm \cite{bib:2}.
Here we are
given a function $f(a)$ defined on the integers $a$ from 1 to N.  The
function has the
property that it takes the value 1 on just a single element of its domain,
$w$, and it
has the value 0 for all $a\neq w$.  With only the ability to call the
function $f$, the task
is to find $w$.  On a classical computer this requires, on average, $N/2$
calls of the
function $f$.  However Grover showed that with a quantum computer $w$ can
be found
with of order $N^{1/2}$ function calls.  This remarkable speed-up
illustrates the power
of quantum computation.  (In the appendix we explain how the Grover
algorithm works.)

In this paper we consider quantum computation differently, as controlled
Hamiltonian
time evolution of a system, obeying the Schrodinger equation
\begin{equation}
i \frac{d}{dt} \ |\psi\rangle = H(t) |\psi\rangle ,
\label{eq:1}
\end{equation}
which is designed to solve a specified problem.  We illustrate this with an
example.
Suppose we are given a Hamiltonian in an N dimensional vector space and we
are told
that the Hamiltonian has one eigenvalue  $E\neq 0$ and all the others are
0.  The
task is to find the eigenvector $| w\rangle$ which has eigenvalue $E$.  We
now give a
solution to this problem and then explain in what sense it is optimal.

We are given
\begin{equation}
H_w = E | w\rangle \langle w|
\label{eq:2}
\end{equation}
with $| w\rangle$ unspecified and $\langle w|w\rangle=1$.  Pick some normalized
vector $|s\rangle$ which of course does not depend on $| w\rangle$ since we
don't yet
know what $| w\rangle$ is.  Now add to $H_w$ the ``driving" Hamiltonian
\begin{equation}
H_D = E|s\rangle \langle s|
\label{eq:3}
\end{equation}
so that the full Hamiltonian is
\begin{equation}
H=H_w + H_D.
\label{eq:4}
\end{equation}
We now calculate the time evolution of the state $|\psi_w,t\rangle$ which
at $t=0$ is
$|s\rangle$,
\begin{equation}
|\psi_w,t\rangle = e^{-iHt} \ |s\rangle.
\label{eq:5}
\end{equation}

It suffices to confine our attention to the two dimensional subspace spanned by
$|s\rangle$ and $|w\rangle$.  The vectors $|s\rangle$ and $|w\rangle$ are
(generally) not orthogonal and we call their inner product $x$,
\begin{equation}
\langle s|w\rangle = x
\label{eq:6}
\end{equation}
where $x$ can be taken to be real and positive since any phase in $\langle
s|w\rangle$
can ultimately be absorbed in  $|s\rangle$.  We will discuss the
expected size of
$x$ shortly.
Now the vectors
\begin{equation}
|r\rangle = \frac{1}{\sqrt{1-x^2}} \left( |s\rangle - x | w\rangle\right)
\label{(eq:7)}
\end{equation}
and $|w\rangle$ are orthonormal.  In the $|w\rangle$, $| r\rangle$ basis the
Hamiltonian (\ref{eq:4}) is
\begin{equation}
H = E \left[\begin{array}{cc} 1 + x^2 & x\sqrt{1-x^2} \\ x\sqrt{1-x^2} & 1-x^2
\end{array}\right]
\label{eq:8}
\end{equation}
and
\begin{equation}
|s\rangle = \left[\begin{array}{c} x \\ \sqrt{1-x^2} \end{array}\right]
\label{eq:9}
\end{equation}
Now a simple calculation gives
\begin{equation}
|\psi_w,t\rangle = e^{-iEt} \left[\begin{array}{c} x \ \cos (E  x  t)  - i
\ \sin (E  x t)
\\
\sqrt{1-x^2}  \cos (E  x  t) \end{array}\right] .
\label{eq:10}
\end{equation}
Thus we see that at time $t$ the probability of finding the state $|w\rangle$ is
\begin{equation}
P(t)=\sin^2  (E  x  t) + x^2 \ \cos^2  (E  x  t)
\label{eq:11}
\end{equation}
and that at a time $t_m$ given by
\begin{equation}
t_m = \frac{\pi}{2Ex}
\label{eq:12}
\end{equation}
the probability is one.

How big do we expect $x$ to be?  In an $N$ dimensional complex vector
space, if you
pick two normalized vectors at random (uniformly on the 2N-1 dimensional unit
sphere), then the expected value of the inner product squared is $1/N$ so
we know that
the expected value of $x$ is of order
$N^{-1/2}$.  Thus starting with
$|s\rangle$, for the probability of finding
$|w\rangle$ to be appreciable we must wait a time of order $N^{1/2}/E$.
This is the
analog analogue of the Grover algorithm result.

Note that the eigenvalues of the Hamiltonian (\ref{eq:8}) are $E(1\pm x)$.
Thus the
difference in eigenvalues is $(2  x  E)$ which is of order $E/N^{1/2}$.  By the
time-energy uncertainty principle, the time required to evolve
substantially, that is
from $|s\rangle$ to $|w\rangle$, must be of order $N^{1/2}/E$ which is the
time we
found.  You might think that by increasing the energy difference, that is
for example, by
using $H_D = E' |s\rangle \langle s|$ with $E'\gg E$ you could speed up the
procedure for
finding $|w\rangle$.  However the next result shows that this is not the case.

We now show that our procedure for finding $|w\rangle$, in a time which
grows like
$N^{1/2}/E$, is optimally short.  The proof we give here is the analog
analogue of
the oracle proof \cite{bib:3} which can be used to show that the Grover
algorithm is
optimal for the problem it sets out to solve.  Again we are given the
Hamiltonian $H_w =
E|w\rangle
\langle w|$ and we wish to add some Hamiltonian $H_D(t)$ to it which drives
the system
to a state which allows us to determine $|w\rangle$.  In an $N$ dimensional
vector
space, there are $N$ linearly independent choices for $|w\rangle$.  We can
pick these to
be a basis for the vector space and we then have
\begin{equation}
\sum_w \ H_w = E\sum_w |w\rangle \langle w| = E .
\label{eq:13}
\end{equation}

The idea of the proof is this:  Start with some initial
$|w\rangle$-independent state
$|i\rangle$ and evolve it with the Hamiltonian
\begin{equation}
H = H_w + H_D (t) .
\label{eq:14}
\end{equation}
After a time $t$ the state we get must be substantially different from what
we would
have gotten using $H_{w\prime} + H_D(t)$ or else we can not  tell
$|w\rangle$ from $|w'\rangle$.  Let
\begin{equation}
i \frac{d}{dt} |\psi_w,t\rangle \  = \ (H_w + H_D (t))|\psi_w,t\rangle
\label{eq:15}
\end{equation}
with
$$|\psi_w,0\rangle=|i\rangle .$$
In order for $|\psi_w,t\rangle$ to differ sufficiently from
$|\psi_{w\prime},t\rangle$ it is
certainly necessary that, for all (but one) $w$,  $|\psi_w,t\rangle$
differs sufficiently
from any
$|w\rangle$-independent vector.  (If some of the $|\psi_w,t\rangle$ were
very close to a
particular $| w\rangle$-independent vector, we could not tell them apart.)  Let
$|\psi, t\rangle$ evolve with $H_D(t)$, that is,
\begin{equation}
i \frac{d}{dt} |\psi, t\rangle \ = \ H_D(t) |\psi,t\rangle
\label{eq:16}
\end{equation}
with
$$|\psi,0\rangle = |i\rangle .$$
We will use $|\psi,t\rangle$ as a $|w\rangle$-independent vector which the
$|\psi_w,t\rangle$ must differ from.
We require $t$ to be large enough that $\Bigl\| |\psi_w,
t\rangle - |\psi,t\rangle\Bigr\|^2 \geq
\varepsilon$ for some fixed $\varepsilon$ which implies
\begin{equation}
\sum_{w} \Bigl\|  |\psi_w,t\rangle - |\psi,t\rangle\Bigr\|^2 \geq N\varepsilon .
\label{eq:17}
\end{equation}
Now consider
\begin{equation}
\frac{d}{dt} \Bigl\|  |\psi_w,t\rangle - |\psi,t\rangle\Bigr\|^2 = -2
\mbox{Re} \
\frac{d}{dt}
\langle
\psi_w,t|\psi,t\rangle
\label{eq:18}
\end{equation}
which upon using (\ref{eq:15}) and (\ref{eq:16}) gives
\begin{eqnarray}
\frac{d}{dt} \Bigl\|  |\psi_w,t\rangle - |\psi,t\rangle\Bigr\|^2 & = & 2 \
\mbox{Im}
\langle\psi_w,t|H_w|\psi,t\rangle \nonumber \\
& \leq & 2 | \langle \psi_w,t|H_w|\psi,t\rangle | \\
& \leq & 2 \Bigl\| H_w |\psi,t\rangle\Bigr\| . \nonumber
\end{eqnarray}
We now sum on $w$ and use the fact that if $\sum^{N}_{i=1} |a_i|^2=1$ then
$\sum^{N}_{i=1} |a_i|\leq N^{1/2}$ along with (\ref{eq:13}) to obtain
\begin{equation}
\frac{d}{dt} \sum_w \Bigl\|  |\psi_w,t\rangle - |\psi,t\rangle \Bigr\|^2
\leq \ 2
EN^{1/2} .
\label{eq:20}
\end{equation}
Since $|\psi_w,0\rangle = |\psi,0\rangle$ we have
\begin{equation}
\sum_w \Bigl\| |\psi_w,t\rangle - |\psi, t\rangle \Bigr\|^2 \leq \ 2
EN^{1/2} t .
\label{eq:21}
\end{equation}
Therefore in order to satisfy (\ref{eq:17}) we must have
\begin{equation}
t \geq \frac{\varepsilon N^{1/2}}{2E} .
\label{eq:22}
\end{equation}
This shows that the $H_D$ we have chosen allows us to determine $|w\rangle$ as
quickly as possible in terms of $N$.

\noindent
\underline{Appendix}: The Grover Algorithm

We are given a function $f(a)$ with $a=1,\ldots N$ such that $f(w)=1$ and
$f(a)=0$ for
$a\neq w$.  We assume that the function $f(a)$ can be calculated using ordinary
(reversible) computer code.  The goal is to find $w$.  Classically this
requires, on
average, $N/2$ evaluations of the function $f$.

We now explain how the Grover algorithm solves this problem; see also
\cite{bib:4}.   The
quantum computer acts on a vector space which has an orthonormal basis
$|a\rangle$ with $a=1, \ldots N$.  It is possible to write a quantum
computer algorithm
which implements the unitary transformation
$$U_f |a\rangle = (-1)^{f(a)} |a\rangle . \eqno (A1)$$
Equivalently we can write
$$U_f = 1-2|w\rangle \langle w| \eqno (A2)$$
The quantum computer algorithm which implements $U_f$ requires two
evaluations of
the function $f$ because it is necessary to erase certain work bits which
we have
supressed.  It is also assumed that the ordinary code which is used to
evaluate $f$ has a
length which does not grow like $N$ to a positive power.  Then the number
of two bit
quantum computer steps required to evaluate $f$ will also not grow as fast
as $N$ to a
power.

Now consider the vector
$$|s\rangle = \frac{1}{N^{1/2}} \sum_a |a\rangle . \eqno (A3)$$
It is also possible to write quantum computer code which implements the unitary
operator
$$U_s = 2|s\rangle \langle s|-1 . \eqno (A4)$$
The number of two bit operations required to implement $U_s$ grows more
slowly than
$N$ to any positive power.

The Grover algorithm consists of letting the operator $U_s
U_f$ act
$k$ times on the vector $|s\rangle$.  To see what happens we can restrict
our attention to
the two dimensional subspace spanned by $|s\rangle$ and $|w\rangle$.  Let
$$|r\rangle = \frac{1}{\sqrt{N-1}} \sum_{a\neq w} |a\rangle \eqno (A5)$$
so that $|w\rangle$ and $|r\rangle$ form an orthonormal basis for the relevant
subspace.  In the $| w\rangle$, $|r\rangle$ basis the operator $U_sU_f$
takes the form
$$U_sU_f = \left[\begin{array}{cc} \cos\theta & -\sin\theta \\ \sin\theta &
\cos\theta
\end{array}\right] \eqno (A6)$$
where $\cos\theta = 1-2/N$.  This implies that
$$(U_sU_f)^k = \left[\begin{array}{cc} \cos(k\theta) & -\sin(k\theta) \\
\sin(k\theta) &
\cos(k\theta)\end{array}\right] . \eqno (A7)$$
Now for $N$ large $\theta \sim 2N^{-1/2}$ so each application of $U_sU_f$
is a rotation
by an angle $\sim 2 N^{-1/2}$.  In the $|w\rangle$, $|r\rangle$ basis, the
initial state
$|s\rangle$ is
$$|s\rangle =\left[\begin{array}{c}N^{-1/2} \\
(1-\frac{1}{N})^{1/2}\end{array}\right] \eqno (A8)$$
which is very close to $|r\rangle$.  However after $k$ steps where
$k\theta=\pi/2$ the
algorithm has rotated the initial state to lie (almost) along $|w\rangle$.
This requires
$k\sim\pi N^{1/2}/4$ steps. Each step actually requires two evaluations of
$f$ so the
number of evaluations of $f$ required to find $w$ grows like $N^{1/2}$.
Accordingly
the number of two bit operations required to implement the algorithm also
grows like
$N^{1/2}$.

\noindent
\underline{Acknowledgement}

E.~F.~ would like to thank the theory group at Universit\`{a} di Roma 1 for
their hospitality and discussions as well as the INFN for partial support.

\end{document}